\documentclass[fleqn,usenatbib]{mnras}

\usepackage{newtxtext,newtxmath}
\usepackage{upgreek}
\usepackage[T1]{fontenc}

\DeclareRobustCommand{\VAN}[3]{#2}
\let\VANthebibliography\thebibliography
\def\thebibliography{\DeclareRobustCommand{\VAN}[3]{##3}\VANthebibliography}

\usepackage{siunitx}
\newcommand{\mps}[0]{\si{\metre\per\second}}
\usepackage{graphicx}	% Including figure files
\usepackage{amsmath}	% Advanced maths commands
\usepackage{xcolor}

\newcommand{\edit}[1]{#1}
\usepackage[colorinlistoftodos,prependcaption,textsize=tiny]{todonotes}

\usepackage{hyperref}
\usepackage{academicons}

\newcommand{\orcid}[1]{\href{https://orcid.org/#1}{\textcolor[HTML]{A6CE39}{\aiOrcid}}}

\usepackage{natbib}

\title[Solar RVs from inactive regions]{The magnetically quiet solar surface dominates HARPS-N solar RVs during low activity}

\author[B. S. Lakeland et al.]{Ben S. Lakeland,$^{1}$ Tim Naylor,$^1$ Rapha{\"e}lle D. Haywood,$^{1}$ Nad{\`e}ge Meunier,$^{2}$ Federica Rescigno,$^{1}$
\newauthor
Shweta Dalal,$^{1}$ Annelies Mortier,$^3$ Samantha J. Thompson,$^4$
Andrew Collier Cameron,$^5$ Xavier Dumusque,$^6$
\newauthor 
Mercedes L\'opez-Morales,$^7$ Francesco Pepe,$^6$  Ken Rice,$^{8,9}$ Alessandro Sozzetti,$^{10}$ St\'ephane Udry,$^{6}$ 
\newauthor  
Eric Ford,$^{11, 12, 13, 14}$ Adriano Ghedina,$^{15}$ and Marcello Lodi$^{15}$
\\
$^{1}$Department of Physics and Astronomy, University of Exeter, Exeter, EX4 4QL, UK\\
$^2$Univ. Grenoble Alpes, CNRS, IPAG, 38000 Grenoble, France\\
$^3$School of Physics \& Astronomy, University of Birmingham, Edgbaston, Birmingham, B15 2TT, UK\\
$^4$Cavendish Laboratory, JJ Thomson Avenue, Cambridge CB3 0HE, UK,\\
$^5$Centre for Exoplanet Science / SUPA, School of Physics \& Astronomy, University of St Andrews, North Haugh ST ANDREWS, Fife, KY16 9SS, UK\\
$^6$Département d’astronomie de l'Université de Genève, Chemin Pegasi 51, 1290 Versoix, Switzerland\\
$^7$Center for Astrophysics ${\rm \mid}$ Harvard {\rm \&} Smithsonian, 60 Garden St, Cambridge, MA 02138, USA\\
$^8$Institute for Astronomy, University of Edinburgh, Royal Observatory, Blackford Hill, Edinburgh, EH9 3HJ, UK\\
$^9$Centre for Exoplanet Science, University of Edinburgh, Edinburgh, EH9 3HJ, UK\\
$^{10}$INAF - Osservatorio Astrofisico di Torino, Via Osservatorio 20, 10025 Pino Torinese, Italy\\
$^{11}$Department of Astronomy \& Astrophysics, 525 Davey Laboratory, Penn State, University Park, PA, 16802, USA\\
$^{12}$Center for Exoplanets and Habitable Worlds, 525 Davey Laboratory, Penn State, University Park, PA, 16802, USA\\
$^{13}$Institute for Computational and Data Sciences, Penn State, University Park, PA, 16802, USA\\
$^{14}$Center for Astrostatistics, 525 Davey Laboratory, Penn State, University Park, PA, 16802, USA\\
$^{15}$Fundaci\'on Galileo Galilei - INAF - Fundaci\'on Canaria, Rambla J.A.Fern\'andez Perez, 7 38712 B.Baja (S.C.Tenerife) Spain
}

\date{Accepted XXX. Received YYY; in original form ZZZ}

\pubyear{2023}

\begin{document}
%\listoftodos[Notes]
\newpage
\label{firstpage}
\pagerange{\pageref{firstpage}--\pageref{lastpage}}
\maketitle

% Abstract of the paper
\begin{abstract}
% 166 words
Using images from the Helioseismic and Magnetic Imager aboard the \textit{Solar Dynamics Observatory} (SDO/HMI), we extract the radial-velocity (RV) signal arising from the suppression of convective blue-shift and from bright faculae and dark sunspots transiting the rotating solar disc. We remove these rotationally modulated magnetic-activity contributions from simultaneous radial velocities observed by the HARPS-N solar feed to produce a radial-velocity time series arising from the magnetically quiet solar surface (the \lq inactive-region radial velocities\rq). We find that the level of variability in the inactive-region radial velocities remains constant over the almost 7 year baseline and shows no correlation with well-known activity indicators. With an RMS of roughly 1 \mps, the inactive-region radial-velocity time series dominates the total RV variability budget during the decline of solar cycle 24.
% The roughly 1 \mps variability shown by the inactive-region RVs is, for large parts of solar cycle 24, significantly larger than the variability in the \lq magnetic activity\rq\  RVs we calculate from the SDO/HMI images. 
Finally, we compare the variability amplitude and timescale of the inactive-region radial velocities with simulations of supergranulation. We find consistency between the inactive-region radial-velocity and simulated time series, indicating that supergranulation is a significant contribution to the overall solar radial velocity variability, and may be the main source of variability towards solar minimum. This work highlights supergranulation as a key barrier to detecting Earth twins.

\end{abstract}

% Select between one and six entries from the list of approved keywords.
% Don't make up new ones.

%%%%%%%%%%%%%%%%%%%%%%%%%%%%%%%%%%%%%%%%%%%%%%%%%%

%%%%%%%%%%%%%%%%% BODY OF PAPER %%%%%%%%%%%%%%%%%%

\section{Introduction}
The radial-velocity (RV) method is one of the most valuable tools in the planet hunters' arsenal, as both a detection method, with over 1000 confirmed exoplanets to date, and a vital follow-up tool for transit missions such as the \textit{Kepler} \citep{2010Sci...327..977B} , \textit{Transiting Exoplanet Survey Satellite} \citep[TESS;][]{2014SPIE.9143E..20R}, and upcoming  \textit{PLAnetary Transits and Oscillations of stars } \citep[PLATO;][]{rauerPLATOMission2014} missions.
Precise RV measurements are often necessary to independently confirm planetary candidates and to determine planetary masses. 
As instrumental precision improves, the intrinsic RV variability of the host star becomes the primary obstacle to detecting and characterising low-mass planets \citep[see][and references therein]{2021arXiv210714291C}.
Stellar variability can mimic non-existent planets \citep{2016MNRAS.456L...6R}, lead to inaccurate mass measurements \citep[e.g., ][]{2023arXiv230608145B, 2023arXiv230614015M}, or completely obscure the signal of a low-mass planet.
\\

Photospheric magnetic activity primarily causes RV variations via two processes. 
Firstly, strong magnetic fields act to inhibit convective blueshift of rising stellar material, imparting a net redshift onto the overall stellar RVs \citep{2010A&A...512A..39M, dumusqueSOAPToolEstimate2014}.
Secondly, bright and dark active regions break the axial symmetry of the rotating stellar disc.
This results in an imbalance of flux from the approaching and retreating stellar limbs, causing a net Doppler shift \citep{1997ApJ...485..319S}. 
As these phenomena occur in isolated, magnetically active regions which can persist for multiple rotation cycles, the resulting RV variability typically displays quasi-periodic behaviour with amplitudes of several \mps.\\

% A number of techniques have successfully been used to account for the RV effects of photospheric magnetic activity.
Traditional activity-mitigation techniques include the use of magnetically sensitive activity indicators which correlate with activity-induced RVs \citep[e.g.,][]{2011A&A...528A...4B, haywoodUnsignedMagneticFlux2022}, modelling RV variations with Gaussian processes to leverage the quasi-periodicity of isolated active regions rotating with the stellar surface \citep[e.g.,][]{haywoodPlanetsStellarActivity2014, 2015MNRAS.452.2269R}, and using simultaneous photometry and ancilliary time series to estimate the effect of active regions \citep[e.g.,][]{aigrainSimpleMethodEstimate2012, barraganPYANETIIIMultidimensional2022}.\\

Whilst these techniques have proven  effective in mitigating the RV variability originating in isolated, magnetically active stellar regions, they do not take into account variability originating from the magnetically quiet stellar surface.
Nevertheless, there are a number of phenomena on the quiet stellar surface which can cause RV variability.
In particular, observations and simulations of solar convective flows (e.g., granulation and supergranulation) have shown RV variability on the order of 0.5-1 \mps \citep[e.g., ][]{meunierUsingSunEstimate2015, meunierUnexpectedlyStrongEffect2019, meunierEffectsGranulationSupergranulation2020, moullaStellarSignalComponents2022}. 
In this paper, we use spatially resolved solar images to calculate the RV impact of isolated magnetically active regions \citep[following][]{2010A&A...519A..66M, milbourneHARPSNSolarRVs2019}.
We remove these contributions from Sun-as-a-star RVs from the HARPS-N spectrograph to isolate and directly characterise the RV impact of the magnetically quiet solar surface (hereafter the \lq inactive-region RVs\rq).\\
% In this paper, we use simultaneous spatially-resolved solar images and Sun-as-a-star RV measurements to isolate and characterise the RV impact of the magnetically quiet solar surface (hereafter the \lq inactive-region RVs\rq).\\

This paper is structured as follows.
In Section \ref{sec:data} we describe the data used in this work and  we isolate the inactive-region RVs in Section \ref{sec:quiet_sun_rvs}.
We analyse the statistical properties of the inactive-region RVs in Section \ref{sec:analysis} and demonstrate that the inactive-region RVs are not simply an instrumental artefact (Section \ref{sec:neid_comparison}).
It is worth noting that, up until this point, our analysis is purely statistical and therefore independent of the physical cause of inactive-region RVs.
In Section \ref{sec:sgr_as_rv_cause}, we show that the inactive-region RVs match simulations of oscillation, granulation, and supergranulation and propose supergranulation as the dominant source of variability.
We conclude in Section \ref{sec:conclusions}.

\section{Data} \label{sec:data}

\subsection{HARPS-N Solar telescope} 

% The HARPS-N Solar telescope has been operating continuously in good weather since the summer of 2015. 
The solar telescope at the Telescopio Nazionale Galileo is a 7.6-cm achromatic lens which feeds the sunlight to an integrating sphere and then into the High Accuracy Radial velocity Planet Searcher for the Northern hemisphere (HARPS-N) spectrograph \citep{dumusqueHARPSNOBSERVESSUN2015,phillipsAstrocombCalibratedSolar2016, colliercameronThreeYearsSunasastar2019, dumusqueThreeYearsHARPSN2021}. 
Observations are taken almost continuously during the day time, with 5-min integration times in order to average over the solar p-mode variations.
% In this data set, to better emulate night-time stellar observations, we average three successive observations to give 15-minute effective integration times. 
RVs are calculated from the extracted HARPS-N spectra using the HARPS-N Data Reduction Software \citep[DRS, ][]{dumusqueThreeYearsHARPSN2021}.
% used for night-time stellar RV measurements. 
By using the same instrument and DRS used for night-time exoplanet searches, the solar feed provides a realistic Sun-as-a-star RV time series and has yielded information about sub-\mps level instrumental systematics present in the HARPS-N spectrograph \citep{colliercameronThreeYearsSunasastar2019}.
To better emulate night-time stellar observations, we average three successive observations to give 15-minute effective integration times. \\

In this paper, we use a data set spanning from 2015-Jul-29 to 2021-Nov-12. The RVs have been transformed into the heliocentric rest frame, allowing the data to be free of planetary signals. Additionally, a daily differential extinction correction was done following the model described in \citet{colliercameronThreeYearsSunasastar2019}. The HARPS-N Solar Telescope operates continuously, observing from under a plexiglass dome. Data affected by clouds or other bad weather thus need to be accounted for afterwards. To remove potential bad data points, we use two metrics for each observation. Firstly, data quality factor $Q$ is calculated from a mixture model approach described in \citet{colliercameronThreeYearsSunasastar2019}. This probability value can be between 0 and 1 where 1 represents the reliable data. Secondly, we extract the maximum and mean count of the exposure meter from the headers of the fits files. The ratio of max and mean, $R$, should be close to 1 for uniform exposures (and is always higher than 1 by definition). A first quality cut is done using the conditions $Q\geq0.99$ and $R<1.5$. From the remaining data points, we fit the histogram of $R$ with a Gaussian function. 
A slight delay in the shutter opening causes this distribution to resemble a Gaussian, rather than the more expected pile-up around 1.
% We note that in our case the shape is Gaussian instead of the perhaps more expected pile-up around 1 due to the shutter opening with a slight delay. 
In a second cut we remove data where $R$ is higher than 3 standard deviations above the mean. Finally, as a last cut, we perform a 5-$\sigma$ clipping on the remaining RVs. The cuts made are very strict and may indeed remove reliable data too, but it is done to ensure the very high quality of data. The remaining data-set has 20~435 observations.

\subsection{SDO/HMI} \label{sec:sdo/hmi}

% From the helioseismic and magenetic imager aboard the \textit{Solar dynamics observatory}  e use 720-second exposures of continuum intensity, line-of-sight Dopplergrams, and 

The Helioseismic and Magnetic Imager aboard the Solar Dynamics Observatory (SDO/HMI) uses six narrow bands around the magnetically-sensitive 6173 \AA\ Fe I line. A Gaussian fit of the narrow-band fluxes as a function of wavelength is used to derive continuum intensitygrams, magnetograms, and line-of-sight velocity maps (Dopplergrams), each spatially resolved with approximately 1" resolution. In this work we use 720-second exposures, every four hours from 2015-Jul-29 to 2021-Nov-12, totalling 12~434 sets of images. We briefly describe how we calculate RVs from these images. For more details, we refer the reader to \citet{haywoodSunPlanethostStar2016} and \citet{milbourneHARPSNSolarRVs2019} which describe in detail the process of extracting disc-averaged physical observables from SDO/HMI images.\\

% \subsubsection{Radial Velocities}

Owing to poor long-term stability of the SDO/HMI instrument, we are not able to directly calculate the solar RVs by simply intensity-weighting the Dopplergram pixels. Fig. 2 of \citet{haywoodUnsignedMagneticFlux2022} shows the effects of long-term systematics when calculating the RVs in this way. There are clear jumps in the RVs caused by instrumental effects. 
This means that we cannot use the Dopplergram to produce long-baseline Sun-as-a-star RV measurements.
% To overcome this instrumental limitation, we use a physically-motivated model to estimate the RV variations arising from two processes on the Sun, in effect using the quiet-Sun pixels as a zero-point calibration to overcome these instrumental effects \citep{haywoodSunPlanethostStar2016, milbourneHARPSNSolarRVs2019, haywoodUnsignedMagneticFlux2022}.\\
To overcome this instrumental limitation, we use a physically-motivated model to estimate the RV variations arising from two processes on the Sun.
In doing so, we calculate the RV variability relative to the quiet Sun \citep{2010A&A...519A..66M, haywoodSunPlanethostStar2016, milbourneHARPSNSolarRVs2019, haywoodUnsignedMagneticFlux2022}.\\

The first of these is a photometric shift, $\delta\mathrm{RV_{phot}}$, caused by a bright plage or dark sunspot breaking the rotational symmetry of the solar disc. In the absence of inhomogeneities on the solar disc, the contribution of the blue-shifted approaching limb and red-shifted retreating limb are in balance, with no overall effect on the solar RVs. If a dark or bright region is present on one of the limbs, the Doppler contribution from that limb is diminished or enhanced, respectively, resulting in an overall RV shift. The second component, $\delta\mathrm{RV_{conv}}$, arises from regions with large magnetic flux suppressing convection on the surface of the Sun. The reduction in the convective blue-shift in these regions imposes a net red-shift onto the overall solar RV. By applying continuum intensity and magnetic flux thresholds \edit{described by \citet{haywoodSunPlanethostStar2016}}, respectively, we isolate the RV contribution of active regions on the Sun.\\
% Both \citet{haywoodSunPlanethostStar2016} and \citet{milbourneHARPSNSolarRVs2019} find that the suppression of convective blue-shift is the dominant effect in the solar RVs.\\
% \begin{figure}
%     \centering
%     \includegraphics[width=\linewidth]{Figures/sdo_instability.pdf}
%     \caption{Full-disc, intensity-weighted RVs of the Sun from \red{dates}, calculated from SDO/HMI images. There are significant jumps in the RV time series due to instrumental effects. This shows that we are unable to directly obtain disc-averaged RVs from SDO/HMI.}
%     \label{fig:sdo_instability}
% \end{figure}

Since SDO/HMI derives its RVs from a single iron line, as opposed to a whole spectrum, \citet{haywoodSunPlanethostStar2016} and \citet{milbourneHARPSNSolarRVs2019} fit a linear combination of these components to solar RVs from HARPS and HARPS-N, respectively, to account for the different scaling of each component.
They found that $\delta\mathrm{RV_{conv}}$ is the dominant of these two effects in the Sun. \citet{milbourneHARPSNSolarRVs2019} only consider the RV contributions from active regions larger than 20 $\upmu$Hem. Those authors find that this cutoff obviates the need to vary the relative contributions of the photometric and convective effects over time. Owing to the fact that the two components forming this RV series are driven by magnetic processes, and that RVs are calculated relative to the quiet Sun, we will refer to the RV series derived from the SDO/HMI images as the \lq magnetic activity\rq\ RVs.
Since the formal uncertainties on the magnetic activity RVs are significantly lower than the uncertainties for the HARPS-N RVs \citep{haywoodUnsignedMagneticFlux2022}, we will refrain from extensive error analysis of the magnetic activity RVs.

\section{Isolating RV contributions from magnetically-inactive regions} \label{sec:quiet_sun_rvs}

In this paper, we are interested in the RV variability originating from magnetically quiet regions of the solar surface. 
To isolate these contributions, we subtract the SDO/HMI RVs (which are only sensitive to large magnetically active regions) from the HARPS-N RVs (which probe the entire solar disc). 
% The resultant time series of residual RVs therefore probes the RV signals from the magnetically quiet solar surface. 
The resultant time series of residual RVs therefore probes the RV signals not arising from the large, magnetically active regions identified by \citet{milbourneHARPSNSolarRVs2019}. 
Throughout this paper, we will refer to this time series as the inactive-region RVs.
To account for the difference in cadence of SDO/HMI and HARPS-N, we interpolate the magnetic activity RVs onto the HARPS-N timestamps.
% % HARPS-N RVs to the magnetic activity RVs derived from SDO/HMI images,
% we interpolate these magnetic activity RVs onto the HARPS-N timestamps, and construct a residual time series by subtracting the SDO/HMI RVs from the HARPS-N RVs. As these RVs probe variability originating from magnetically acive solar regions and the entire solar disc, respectively, the time series of differences
In principle, this could introduce a signal into the inactive-region RVs, since we are smoothing over all variability in the magnetic activity RVs over timescales shorter than 4 hours. Recalling, however, that the magnetic activity RVs encode variability occurring on rotational timescales, and noting that shorter-timescale processes (e.g., oscillations/granular flows) do not contribute to the magnetic activity RVs, we are justified in this approach. We provide a quantitative justification in Appendix \ref{sec:app_interp}.
It is worth highlighting that the model of \citet{milbourneHARPSNSolarRVs2019}, which we use in this paper, does not include the effect of the smallest active regions.
However, those authors show that by only considering the largest active regions, they reproduce the majority of the rotationally-modulated, activity-induced RVs, without having to introduce an arbitrary trend.
The models of \citet{milbourneHARPSNSolarRVs2019} reduce the HARPS-N RV scatter from 1.65 \mps to 1.21 \mps and 1.18 \mps with and without the $20$ $\upmu$Hem active-region cutoff, respectively\footnote{To reduce the RV scatter to 1.18 \mps when taking into account all active regions, an arbitrary linear drift was introduced. Without that drift, the scatter is reduced to 1.31 \mps.}.
We are therefore justified in describing the RV time series produced by removing this contribution from the disc-integrated HARPS-N RVs the \lq inactive-region\rq\ RVs.
We do note however, that there will inevitably remain a small component in the inactive-region RVs that corresponds to the smallest-scale magnetic regions.
\\

% \begin{figure}
%     \centering
%     \includegraphics[width=\linewidth]{Figures/quiet_sun_detrended_new.pdf}
%     \caption{Inactive-region RV time series shown before (top) and after (bottom) de-trending with a 100-day rolling mean.}
%     \label{fig:quiet_sun_rvs}
% \end{figure}

To avoid any effects of long-term instrumental effects, we de-trend the inactive-region RVs with a 100-day rolling mean, following the treatment of \citet{moullaStellarSignalComponents2022}.
%When we took the pure residuals between the HARPS-N and the SDO/HMI RVs, there was a linear trend towards the end of the time series. To remove any effect of the magnetic cycle, or any long-term instrumental drift, we follow \citet{moullaStellarSignalComponents2022} and smooth the quiet Sun RVs with a 100-day rolling mean.
We note that such a de-trending preserves both the amplitude and timescale of any short-term variability present in the RV time series, and so will not affect the results of this work. We opt to smooth the inactive-region RVs, as opposed to the HARPS-N RVs, since we are interested in the difference between the RVs measured by HARPS-N and those derived from SDO/HMI images. 
Directly smoothing the HARPS-N RVs would risk smoothing over the rotationally modulated activity signal present in both series, thereby imposing an artificial signal in the residuals when we subtract the un-smoothed magnetic activity RVs.\\

% To assess the impact of instrumental effects on our results, we use NEID solar RVs (section \red{blah}) to produce a second inactive-region RV time series following the technique laid out in Section \ref{sec:quiet_sun_rvs}.\footnote{\red{Something about the A and B coefficients}} The HARPS-N and NEID RVs overlap from \red{DATES} and we compare the inactive-region RVs during this overlap to investigate the effect of instrumental systematics. In Fig \ref{fig:hn_neid_comparison}, we plot structure functions (see Appendix \ref{ of both the HARPS-N and NEID inactive-region RVs. Fig \ref{fig:hn_neid_comparison} shows that the two time series have very similar variability characteristics, despite the different instruments. The agreement of these time series indicates that instrumental effects do not significantly alter the results of this work and demonstrates that the variability seen in the inactive-region RVs is predominately astrophysical.\\

% \subsection{\red{How do we know this is a quiet Sun time series?}}

% \subsubsection{\red{Correlation plots}}

\begin{figure*}
    \centering
    \includegraphics{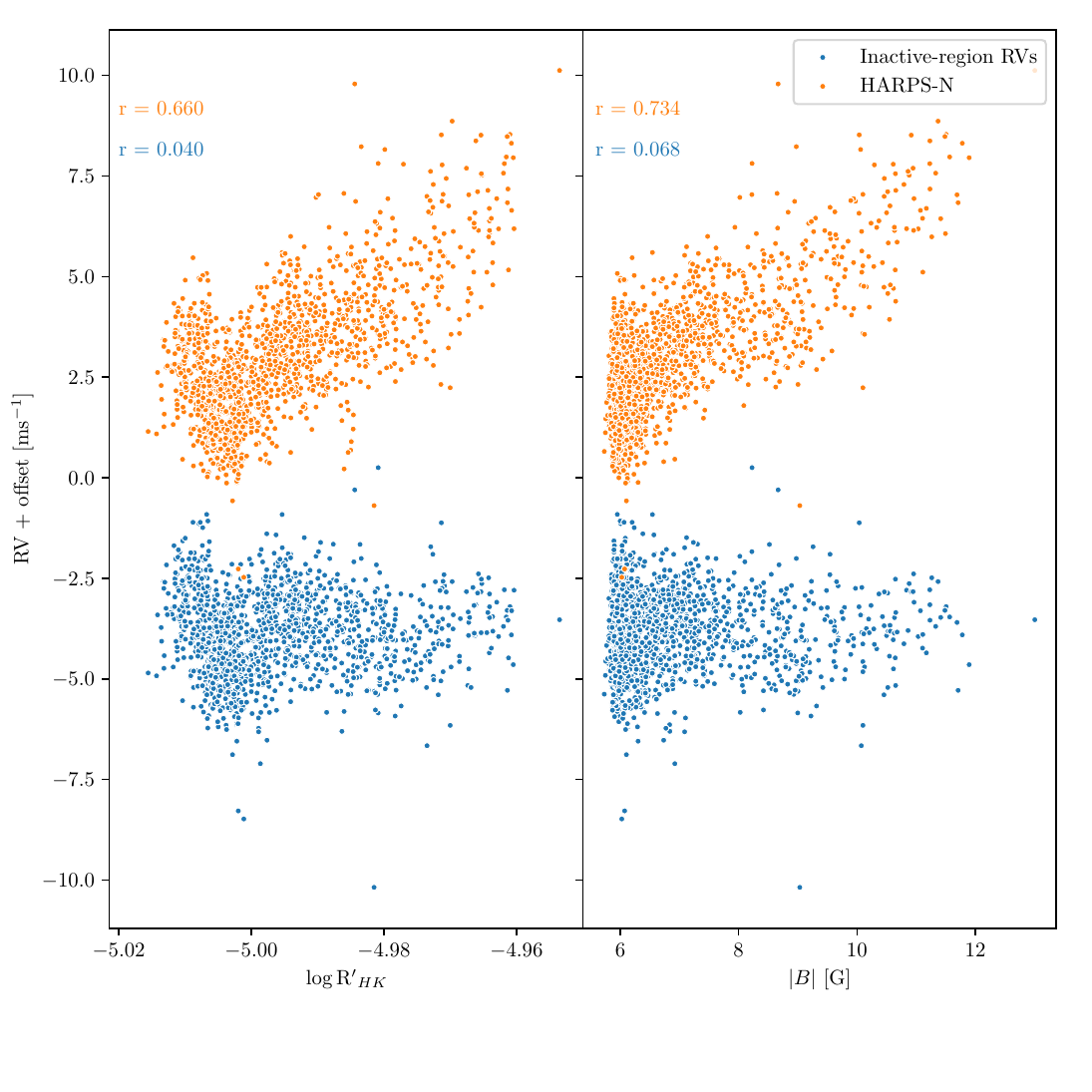}
    \caption{Scatter plots showing the correlation between daily mean RVs (orange for the HARPS-N RVs and blue for the inactive-region RVs) and $\log{R'_\mathrm{HK}}$ (left) and unsigned magnetic flux (right). Both scatter plots demonstrate an almost complete removal of any correlation between the RVs and the respective activity indicator after subtracting the magnetic activity RVs. For each data set, we also numerically show the Pearson correlation coefficient. RV data are shown with an offset for clarity.}
    \label{fig:correlations}
\end{figure*}

\begin{figure*}
    \centering 
    \includegraphics[width=\linewidth]{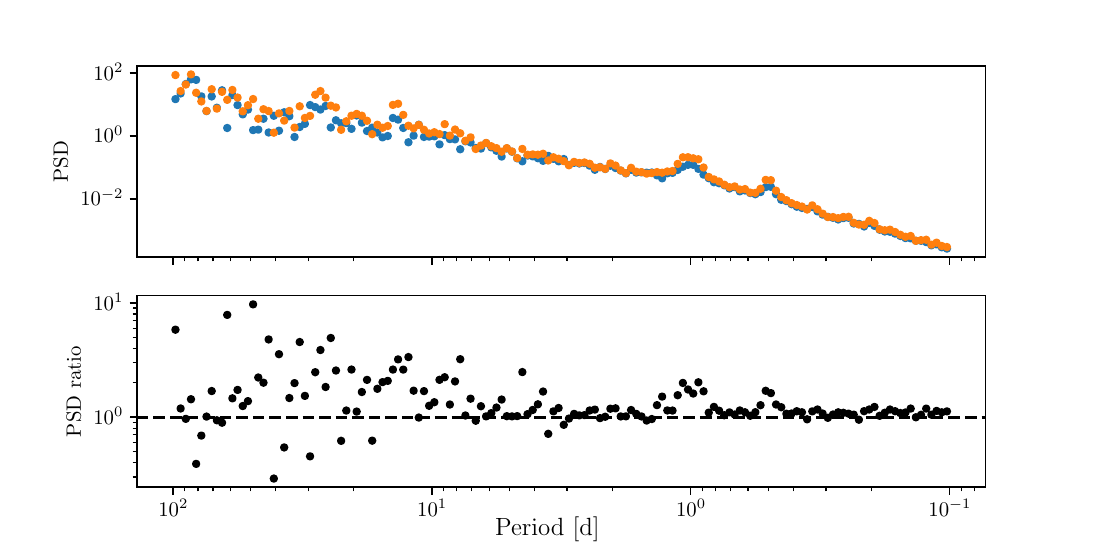}
    \caption{(Top) The power spectral density (PSD) of HARPS-N RVs (orange) and of the inactive-region RVs (blue), calculated in 150 logarithmically-spaced bins in frequency. (Bottom) The ratio of HARPS-N RV PSD to inactive-region RV PSD. Although there is still a noticeable feature in the residual RV PSD at the rotation period, and at one day and its alias, we see lower power at timescales longer than $\sim10$ days when compared to the full HARPS-N RVs. By contrast, the two PSDs are very similar at shorter timescales. This indicates we are successful in removing much of the variability at timescales around or above the solar rotation period, but we leave the shorter timescale variability largely unaffected.}
    \label{fig:psd_plot}
\end{figure*}

Fig. \ref{fig:correlations} shows the correlation between the daily mean RV and both $\log \mathrm{R'}_{HK}$ and unsigned magnetic flux \citep{haywoodUnsignedMagneticFlux2022} values for the HARPS-N RVs and the inactive-region RVs, as well as the respective Pearson correlation coefficients. We see an almost total removal of any correlation between the RVs and the activity indices when the magnetic activity RVs are subtracted from the HARPS-N RVs, indicating that we are successfully removing the majority of the RV signal caused by magnetic activity.\\

In addition, following \citet{moullaStellarSignalComponents2022}, we calculate the velocity power spectrum (PS) of the HARPS-N RVs and the residual RVs as follows. Each RV time series can be represented as a linear combination of sinusoidal components,
\begin{equation}
    \mathrm{RV}(t) = c + \sum_{\nu} a(\nu) \cos{(2 \pi \nu t)} + b(\nu) \sin{(2 \pi\nu t)},
\end{equation}
where $a(\nu)$ and $b(\nu)$ are frequency-dependent coefficients, and $c$ is a constant offset. The velocity power spectrum is defined as
\begin{equation}
    \mathrm{PS}(\nu) = a(\nu)^2 + b(\nu)^2,
\end{equation}
and is converted to velocity power spectral densities (PSDs) by multiplying by the effective length of the observing window \citep{2005ApJ...635.1281K}.\\

Fig. \ref{fig:psd_plot} shows the two PSDs as well as the ratio of power in the HARPS-N RVs to the inactive-region RVs, calculated for logarithmically-spaced bins in frequency. We see evidence of decreased power in the inactive-region RVs at timescales comparable to and longer than the solar rotation period, whereas the two PSDs are similar for shorter timescales. This further suggests that we are successful at removing most of the rotationally modulated variability, whilst retaining the short-timescale variability characteristics of the HARPS-N RVs.

\section{Inactive-region RVs exhibit a constant level of variability} \label{sec:analysis}

\begin{figure}
    \centering
    \includegraphics[width=\linewidth]{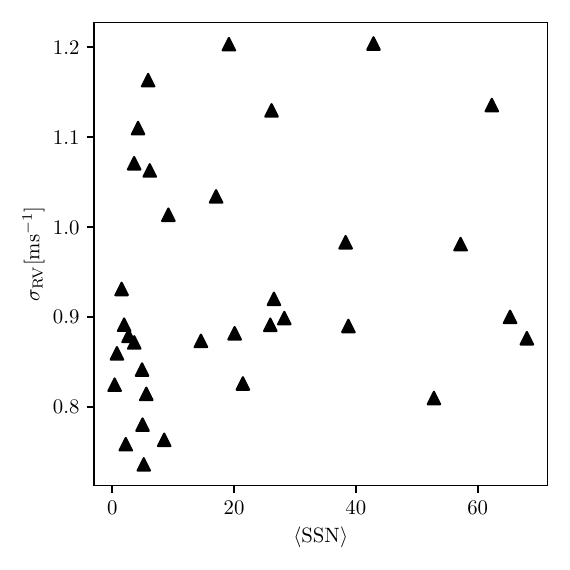}
    \caption{Scatter plot of the mean sunspot number and the standard deviation of inactive-region RV, calculated for non-overlapping 50-day periods from 2015 to 2021. We see no correlation between RV RMS and mean sunspot number.}
    \label{fig:ssn_vs_rms}
\end{figure}

Having established that we are successful in removing the RV signature of large magnetically active regions (Section \ref{sec:quiet_sun_rvs}), we now analyse the inactive-region RVs. To investigate how the inactive regions vary throughout the solar cycle, we compare the RMS of the RVs with the mean sunspot number\edit{, taken from \citet{sidc},} for 45 50-day intervals, as a proxy for overall activity level. In Fig. \ref{fig:ssn_vs_rms} we show that there is no correlation. Additionally, despite the sunspot number changing by more than two orders of magnitude, the standard deviations of the RVs vary only slightly. This shows that the
% for 50-day sections in Fig \ref{fig:ssn_vs_rms}. We see no significant trend between the sunspot number and RMS. This indicates the 
inactive-region RV variability remains relatively constant at a wide range of magnetic-activity levels.
In Appendix \ref{app:sigma_sigma}, we show that the 17 per cent scatter in the RMS shown here is consistent with being dominated by sampling noise.
\\

To investigate how the inactive-region RVs behave at different timescales, we use structure functions \citep[SFs; see e.g,. ][]{1985ApJ...296...46S, sergisonCharacterizingIbandVariability2020,lakelandUnderstandingYSOVariability2022}. Structure functions quantify how the variability within a time series changes with timescale. In Appendix \ref{sec:app_sf_rms} we show that, for a continuous, uncorrelated signal, the structure function is equivalent to 
\begin{equation}
    \mathrm{SF}(\tau) = 2 \sigma^2,
\end{equation}
where $\sigma$ is the RMS of the signal. We therefore choose to use $\sqrt{\frac{1}{2}\mathrm{SF}}$ to quantify the RV variability at a given timescale $\tau$, emphasising the connection with the RMS. We provide a more in-depth description of structure functions and their key properties required for this analysis in Appendix \ref{sec:structure_functions}.\\

\begin{figure}
    \centering
    \includegraphics[width=\linewidth]{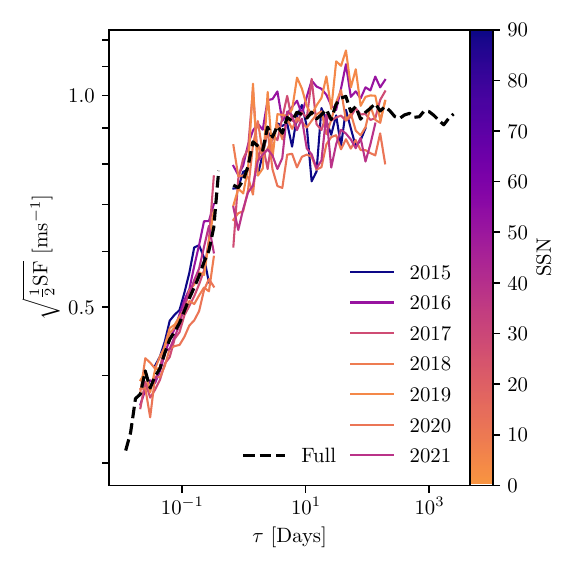}
    \caption{Structure functions of inactive-region RVs, divided by calendar year. 
The lines are coloured by average sunspot number in that year.    
    % The darkest line is for data collected in 2015 (at relatively high solar activity) and the lightest for data collected in 2022 (at low solar activity).
    Despite data collection spanning a wide range of magnetic activity levels, the variability spectrum for the inactive-region RV shows no significant variation from year to year. The black dashed line shows the structure function for the entire inactive-region RV series. We note the gap at $\tau\sim0.5$ d. This is typical of ground-based telescopes, which cannot observe the Sun at night time.}
    \label{fig:structure_functions}
\end{figure}

We now segment the RVs by calendar year and show the corresponding structure functions in Fig. \ref{fig:structure_functions}. 
We opt to use longer sub-series than the 50-day samples in Fig. \ref{fig:ssn_vs_rms} to ensure that each timescale is well sampled, with many pairs of data points contributing to the calculation for each timescale. 
We see that despite covering a wide range of magnetic activity levels, the inactive-region RVs have similar variability at all timescales.
We also note that, for each 1-year segment, there is no significant increase in variability at timescales longer than $\sim$ 3 days.
% We see similar results to Fig \ref{fig:ssn_vs_rms}, with the level of variability remaining constant over the course of the data set. Using structure functions however, we show that the level of variability in the inactive-region RVs remains similar at all timescales
% This roughly constant level of variability at all timescales is in agreement with Fig. \ref{fig:ssn_vs_rms}.
The typical year-to-year variation is 5-10 per cent.
This is comparable to the typical lower-bound sampling error we estimate in Appendix \ref{sec:app_sf_err}, indicating that the true year-to-year variation is very low.
\\

\begin{figure*}
    \centering
    \includegraphics[width=\linewidth]{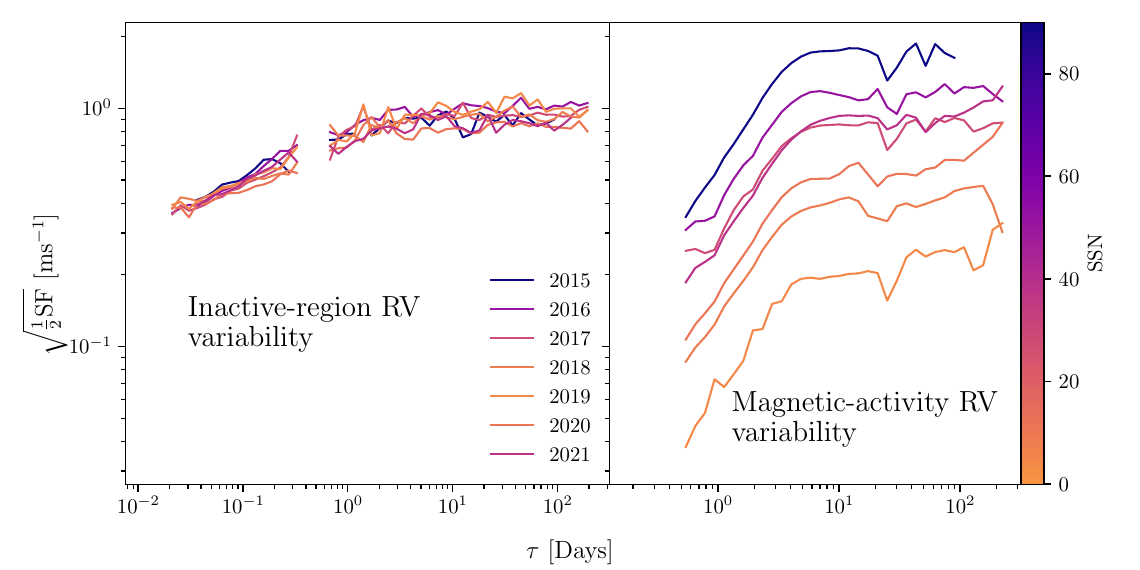}

    \caption{(Left) The structure functions shown in Fig. \ref{fig:structure_functions} of inactive-region RVs. (Right) The structure functions of the corresponding year-long run of magnetic-activity RVs. To illustrate the relative level of variability in the inactive-region and magnetic activity RVs, both sets of structure functions are shown on the same scale. The inactive-region RVs consistently show a level of variability around 1 \mps. The relatively constant level of variability in the inactive-region RVs contrasts drastically to the large range of variability levels demonstrated in the magnetic activity RVs. Additionally, the level of variability shown by the inactive-region RVs is often much larger than that in the magnetic activity RVs. As in Fig. \ref{fig:structure_functions}, both sets of structure functions are coloured by the average sunspot number during the calendar year in which the RVs were obtained.}
    \label{fig:structure_functions_qs_sdo}
\end{figure*}

\begin{figure}
    \centering
    \includegraphics[width=\linewidth]{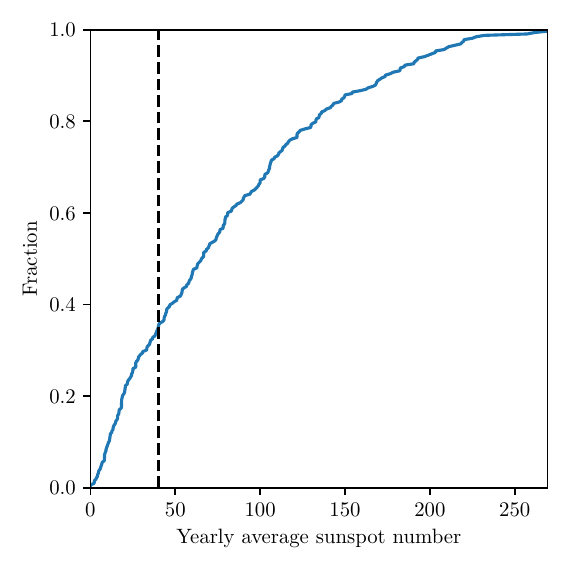}
    \caption{Cumulative distribution of the yearly-average sunspot number, dating from 1700 to present. The average sunspot number between 2016-Jan-01 and 2016-Dec-31 is indicated with a vertical dashed line. Roughly 35 per cent of recorded years have a lower average sunspot number.}
    \label{fig:ssn_ecdf}
\end{figure}

We contrast this with the behaviour of the magnetic-activity RVs over the same baseline.  We calculate structure functions of the magnetic-activity RVs (again segmented by calendar year). We plot structure functions calculated from both the inactive-region RVs and magnetic activity RVs on the same axis scales in Fig. \ref{fig:structure_functions_qs_sdo} to highlight the contrast. Whilst the magnetic-activity RVs vary by more than an order of magnitude, the inactive-region RVs exhibit an almost unchanging level of variability.
Fig. \ref{fig:structure_functions_qs_sdo} also demonstrates that, for the majority of the 7-year baseline, the variability in the residual RVs is comparable to or larger than the variability shown in the magnetic activity RVs. 
This shows that, for the quieter period of solar cycle 24, the majority of the RV variability originated from magnetically-inactive regions, as opposed to magnetically-active regions.
The two contributions are approximately equal during the 2016-2017 period, where the average sunspot number was 39.8.
Fig. \ref{fig:ssn_ecdf} shows the distribution of yearly-averaged sunspot numbers, dating back to 1700 from the World Data Center SILSO, Royal Observatory of Belgium, Brussels \citep{sidc}.
The vertical line corresponds to the 2016-2017 average.
Roughly 35 per cent of recorded years have a lower average sunspot number.
This implies that the inactive-region RVs may dominate solar RVs for roughly one third of the time.
\\

\section{Inactive-region RVs are not dominated by instrumental effects} \label{sec:neid_comparison}

It is in principle plausible that the variability shown in Fig. \ref{fig:structure_functions} at timescales of around two days is due to the instrumental characteristics of the HARPS-N spectrograph, rather than genuine solar effects. To address this concern, we repeat our analysis with RVs from the NEID solar feed \citep{linObservingSunStar2022} taken between 2021-Jan-01 and 2022-Jun-13. 
We use RVs reported by version 1.1 of the NEID Data Reduction Pipeline\footnote{\url{https://neid.ipac.caltech.edu/docs/NEID-DRP/index.html}} and select observations that are free from known issues due to the daily wavelegnth calibration, a known cabling issue or evidence of poor observing conditions based on the pyrheliometer and exposure meter data. 
\edit{We select only observations taken before 17:30 and with airmass less than 2.25 to reduce calibration and atmospheric effects.}
% doi:10.5281/zenodo.7857521
Due to the different wavelength ranges and sensitivities of the NEID and HARPS-N spectrographs, the photometric shift, $\delta \mathrm{RV_{phot}}$, and suppression of convective blueshift, $\delta \mathrm{RV_{conv}}$, (see Section \ref{sec:sdo/hmi}) may have different contributions to the overall RVs. To account for this, following \citet{milbourneHARPSNSolarRVs2019}, we fit a linear combination of these components to the NEID RVs.\\

\begin{figure}
    \centering
    \includegraphics[width=\linewidth]{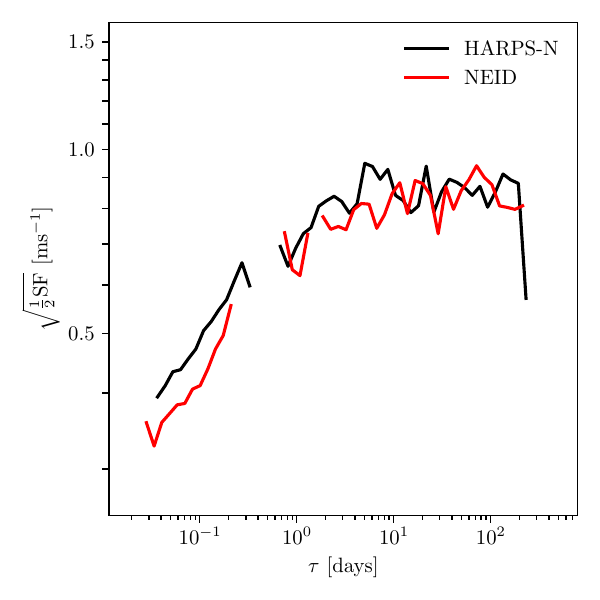}
    \caption{Structure function comparison of inactive-region RVs calculated by subtracting the RV signatures of magnetically-active regions (Section \ref{sec:sdo/hmi}) from HARPS-N (black) and NEID (red) RVs between 2021-Apr-01 and 2021-Nov-12. Both structure functions demonstrate similar variability characteristics despite the different spectrographs, indicating that the variation seen in the inactive-region RVs is predominately astrophysical in nature, not instrumental.}
    \label{fig:hn_neid_comparison}
\end{figure}

It is worth noting that significant intra-day systematics appear in the NEID solar RVs due to, amongst other effects, unaccounted-for differential extinction. 
% We correct for the sub-day variations by removing an empirical cubic trend. 
We model the sub-day variations with a best-fit cubic trend to the 1-day phase-folded RVs, which we then subtract from the NEID solar RVs. 
Whilst this is not as sophisticated as the treatment of \citet{colliercameronThreeYearsSunasastar2019}, we wish simply to recover the general variability properties seen in Fig. \ref{fig:structure_functions} and so a full treatment of NEID instrumental effects is both beyond the scope of this paper and not necessary for our purposes. As with the HARPS-N RVs, we use the recalculated SDO/HMI RVs and NEID RVs to produce a time series of inactive-region RVs. The HARPS-N and NEID RVs temporally overlap between 2021-Apr-01 and 2021-Nov-12 and we compare the HARPS-N- and NEID-derived inactive-region RV time series during this overlap to investigate the effects of instrumental systematics. Fig. \ref{fig:hn_neid_comparison} shows structure functions of the two inactive-region RV series. 
We find that the variability of these two time series at all timescales typically differs by less than 10 per cent, comparable to the lower-bound sampling uncertainty we estimate in Appendix \ref{sec:app_sf_err}.
% We find good agreement between the two structure functions, 
This indicates that the results of section \ref{sec:analysis} are genuine, and are not simply instrumental artefacts, and that any effects of instrumental differences are smaller than the intrinsic variability of the solar RVs; as such, we will focus the remainder of our analysis on the longer-baseline time series of inactive-region RVs derived from HARPS-N.
We note that this analysis qualitatively agrees with the results of \citet{2023AJ....166..173Z}, who compare solar RVs from state-of-the-art spectrographs, including HARPS-N and NEID. 
Those authors find remarkable agreement between the different instruments.

\section{Supergranulation as the driver of inactive-region RVs}\label{sec:sgr_as_rv_cause}

In Section \ref{sec:analysis}, we demonstrated that the inactive-region RVs have a roughly constant level of variability over the decline of solar cycle 24. To determine the cause of this variability, we compare the inactive-region RVs to a simulation of the oscillation, granulation, and supergranulation of a G2 star from \citet{meunierUnexpectedlyStrongEffect2019}, based on the technique of \citet{harvey1984probing} and the results of  \citet{meunierUsingSunEstimate2015}.
The simulated RV time series is averaged over 15 minutes to best match the HARPS-N observations.
\\

% We therefore take the latter \red{level of variability} as the level of granulation present in the quiet Sun RVs.\\

\begin{figure}
    \centering
    \includegraphics[width=\linewidth]{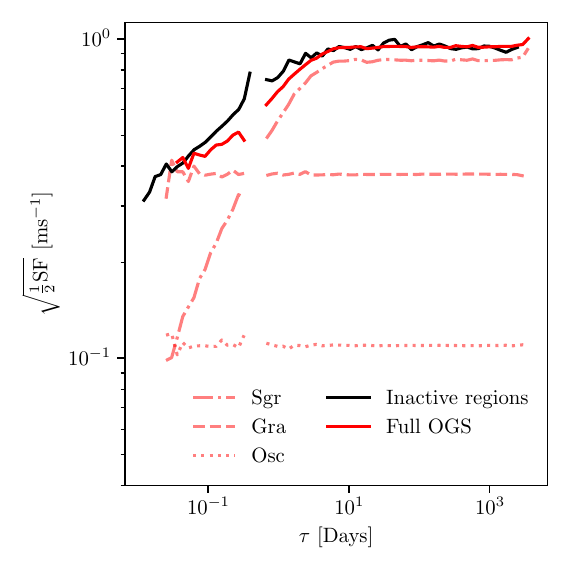}
    
    \caption{Structure function comparison of the inactive-region RVs (black), simulated oscillation, granulation, and supergranulation (OGS) RVs (pink), and the combined simulated RV time series (red). The simulated RVs show good agreement with the inactive-region RVs and indicate that supergranulation is a significant source of RV variability in the Sun, especially at timescales longer than a few hours.}
    \label{fig:meunier_sf}
\end{figure}

Fig. \ref{fig:meunier_sf} shows structure functions of each component in the simulation of \citet{meunierActivityTimeSeries2019} (pink), the combined simulated time series (red), and the inactive-region RVs (black). 
The granulation (Gra) and supergranulation (Sgr) components are re-scaled by a constant factor to match the observations; the oscillation (Osc) component is assumed to be negligible \edit{\citep[e.g.,][]{2019AJ....157..163C}} and so is not re-scaled. However, we opt to include the oscillation component for completeness.
Since granulation and supergranulation dominate at different timescales, we are able to re-scale the simulated RVs to constrain the contribution of each process by matching the power in the inactive-region RVs at short and long timescales, respectively.
By combining and re-scaling the simulated RVs in this way, we are able to accurately reproduce both the overall variability amplitude of the inactive-region RVs, and the timescale of the inactive-region RVs.\\

\citet{meunierActivityTimeSeries2019} found that the amplitude for the RV signals from granulation is 0.8 \mps, though there is evidence that an amplitude of 0.4 \mps is a more appropriate level \citep[e.g., ][]{2020A&A...635A.146S, 2023A&A...670A..24S}.
We see in Fig. \ref{fig:meunier_sf} that the choice to use the lower level of granulation variability is justified; a granulation RV time series with RMS of 0.8 \mps would have more power at short timescales than is seen in the true RVs. 
% We find that the optimal combination of granulation and supergranulation components gives granulation and supergranulation variability amplitudes of 0.37 \mps and 0.86 \mps, respectively. 
We find granulation and supergranulation variability amplitudes of 0.37 \mps and 0.86 \mps, respectively.
We therefore find a similar level of granulation to the lower level of \citet{meunierEffectsGranulationSupergranulation2020}, and to the 0.33 \mps prediction of \citet{2023MNRAS.525.3344D}.
We note that the 0.86 \mps level of supergranulation we find is around 25 per cent higher than the 0.68 \mps level found by \citet{moullaStellarSignalComponents2022}, though the granulation amplitudes are similar.
Given the difference in methodology between this work and that of \citet{moullaStellarSignalComponents2022}\edit{, who use asteroseismology techniques to analyse Sun-as-a-star RVs from the HARPS and HARPS-N spectrographs}, it is unsurprising that there is a difference in the exact amplitude of supergranulation derived.
Both analyses demonstrate that supergranulation contributes a significant fraction of the total solar RV variability.\\

Our interpretation that the inactive-region RVs are primarily caused by granulation and supergranulation is supported by the fact that the level of variability is constant at various levels of the solar cycle (Figs. \ref{fig:ssn_vs_rms} \& \ref{fig:structure_functions_qs_sdo}). \citet{mullerDoesSolarGranulation2018} used HINDODE/SOT images of the Sun from 2006-Nov to 2016-Feb to investigate photometric properties of solar granules. Over this period, they did not find changes in either the granulation contrast or granulation scale of the 3 per cent level at which they were sensitive, indicating no significant variation of granulation properties over the solar cycle.
\\

% We note that we have not considered the effect of instrumental systematics in this analysis. 
We note that Fig. \ref{fig:meunier_sf} does not include any instrumental systematics.
Whilst this makes it difficult to evaluate numerical uncertainties on the level of supergranulation quoted above, it is unlikely to significantly change the amplitude.
Firstly, a white noise profile (such as from photon noise) would inject power at all timescales and appear flat as a structure function. 
Since we constrain the granulation amplitude by using the short-timescale power, we can be confident that we have not underestimated the uncorrelated noise.
Thus the re-scaling we apply to the supergranulation time series to match the long-timescale power is reasonably robust against the level of photon noise, which is estimated to be as low as 0.24 \mps \citep{moullaStellarSignalComponents2022}.
Secondly, we show in Section \ref{sec:neid_comparison} that the level of variability of inactive-region RVs derived from HARPS-N and from NEID observations are consistent within typical sampling errors at all timescales, indicating that the variability observed is not dominated by instrument-specific artefacts.
This shows that the supergranulation amplitude derived from Fig. \ref{fig:meunier_sf} is representative.\\

Fig. \ref{fig:meunier_sf} shows that the combined simulated RVs show slightly reduced power on intermediate timescales (between a few hours and a few days) as compared to the inactive-region RVs. 
Unaccounted-for correlated instrumental noise (e.g., an imperfect correction for differential extinction) could inject variability at intermediate timescales to account for this discrepancy. Alternatively, the intrinsic variability spectrum of supergranulation may differ slightly to that of granulation. The simulations of \citet{meunierUsingSunEstimate2015} assume that supergranules evolve similarly to granules. The resulting simulated time series therefore share a common variability power law. A relaxation of this assumption may account for the mismatched power at intermediate timescales.
Additionally, the smallest-scale active regions, unaccounted for in the model of \citet{milbourneHARPSNSolarRVs2019}, may inject RV variability at these intermediate timescales.\\

Regardless of the cause of this discrepancy, the agreement of the inactive-region RVs with both the amplitude and timescale of the simulations of \citet{meunierActivityTimeSeries2019} and analysis of  \citet{moullaStellarSignalComponents2022} offers good evidence that the variability seen in the inactive-region RVs is predominantly caused by supergranulation at timescales of a few days and longer. 
Given that these RVs exhibit a relatively constant level of variability over the solar cycle (Figs. \ref{fig:structure_functions} \& \ref{fig:structure_functions_qs_sdo}), we provide evidence that supergranulation dominates the solar RVs on the approach to cycle 24 minimum and therefore poses a significant barrier to the detection of Earth twins whose Doppler shifts are $\sim 0.1$ \mps .\\

\section{Conclusions}\label{sec:conclusions}

We have generated solar RVs from SDO/HMI images following the technique of \citet{milbourneHARPSNSolarRVs2019}. The two components accounted for in this time series are the RV signatures arising from the suppression of convective blueshift in magnetically active regions and the effect of photometrically imbalanced regions transiting the rotating solar disc. We subtract these \lq magnetic activity\rq\ RVs from the disc-integrated RVs obtained from the HARPS-N solar feed to isolate the RV contribution from the magnetically quiet solar surface.\\

We find that the resulting \lq inactive-region\rq\ RV time series shows no correlation with the well known activity indicators $\log \mathrm{R'}_{HK}$ and unsigned magnetic flux. This, combined with the reduced Fourier power at timescales longer than roughly half a solar rotation period, show we are successful in removing the majority of the RV signal arising from the active regions on the Sun.\\

We show that the inactive-region RV variability is relatively stable, showing no significant change in level with either time or sunspot number. This contrasts starkly with the magnetic-activity RVs, which show a change in variability of more than an order of magnitude over the same baseline. This relative constancy in the inactive-region RV variability means that, when the Sun is relatively quiescent, the solar RVs are dominated by the RV signal originating in inactive regions, not active regions.
We find that the two contributions are roughly even when the average sunspot number is around 40.
This level of magnetic activity represents the 35$^\mathrm{th}$ percentile recorded in archival sunspot data dating from 1700.\\

% Finally, we find that the dominant timescale present in the quiet Sun signal is comparable to timescales of supergranulation, both from helioseismic surveys and analytic models. 

Finally, we compare the inactive-region RV time series to recent models of oscillation, granulation, and supergranulation \citep{meunierUnexpectedlyStrongEffect2019}. We find good agreement with both  variability amplitude and timescale between the inactive-region and simulated RV time series.
We find that  granulation and supergranulation induce RV variability with amplitudes of 0.37  \mps and 0.86 \mps, respectively.
In doing so, we provide empirical evidence that supergranulation can dominate solar RVs and therefore pose a significant barrier to detection of Earth twins. 
This is in agreement with the findings of \citet{2023arXiv230614015M}.

\section*{Acknowledgements}

% \red{Thank adam for quiet-Sun RVs.}
We would like to thank Suzanne Aigrain and Niamh O'Sullivan for assisting in normalising the PSD of Fig. \ref{fig:psd_plot}.
B.S.L. is funded by a Science and Technology Facilities Council (STFC) studentship (ST/V506679/1).
R.D.H. and S.D. are funded by the STFC's Ernest Rutherford Fellowship (grant number ST/V004735/1).
F.R. is funded by the University of Exeter’s College of Engineering, Maths and Physical Sciences, UK. S.J.T. is funded by STFC grant number ST/V000918/1.
A.C.C. acknowledges support from STFC consolidated grant number ST/V000861/1, and UKSA grant number ST/R003203/1.
This project has received funding from the European Research Council (ERC) under the European Union’s Horizon 2020 research and innovation programme (grant agreement SCORE No 851555).
This work has been carried out within the framework of the NCCR PlanetS supported by the Swiss National Science Foundation (SNSF) under grants 51NF40\_182901 and 51NF40\_205606
F.P. would like to acknowledge the SNSF for supporting research with HARPS-N through the SNSF grants 140649, 152721, 166227, 184618 and 215190. The HARPS-N Instrument Project was partially funded through the Swiss ESA-PRODEX Programme.
K.R. acknowledges support from STFC Consolidated grant number ST/V000594/1.
This research was supported by Heising-Simons Foundation Grant \#2019-1177 and NASA Grant \# 80NSSC21K1035 (E.B.F.).This work was supported in part by a grant from the Simons Foundation/SFARI (675601, E.B.F.).
The Center for Exoplanets and Habitable Worlds is supported by Penn State and its Eberly College of Science.
\\

This work is based 
in part
on observations at Kitt Peak National Observatory, NSF’s NOIRLab, managed by the Association of Universities for Research in Astronomy (AURA) under a cooperative agreement with the National Science Foundation. The authors are honoured to be permitted to conduct astronomical research on Iolkam Du\'ag (Kitt Peak), a mountain with particular significance to the Tohono O\'odham.
% The authors express particular gratitude to Michael Palumbo III and Eric Ford for sharing a data set with \red{quality cuts} .
 
Data presented herein were obtained at the WIYN Observatory from telescope time allocated to NN-EXPLORE through the scientific partnership of the National Aeronautics and Space Administration, the National Science Foundation, and the National Optical Astronomy Observatory.

We thank the NEID Queue Observers and WIYN Observing Associates for their skillful execution of our NEID observations. 
In particular, we express gratitude to Michael Palumbo III for compiling the NEID RVs used in this study.\\

%To be used until at least Sep 2023:
% Deepest gratitude is expressed to Zade Arnold, Joe Davis, Michelle Edwards, John Ehret, Tina Juan, Brian Pisarek, Aaron Rowe, Fred Wortman, the Eastern Area Incident Management Team, and all of the firefighters and air support crew who fought the recent Contreras fire who, against great odds, saved Kitt Peak National Observatory.

%%%%%%%%%%%%%%%%%%%%%%%%%%%%%%%%%%%%%%%%%%%%%%%%%%
\section*{Data Availability}

This work is underpinned by the following publicly available datasets: SDO/HMI images, available at \url{https://sdo.gsfc.nasa.gov/data/aiahmi/}; and NEID solar RVs, made available at \url{https://zenodo.org/record/7857521}.
In addition, this work makes extensive use of the HARPS-N solar RVs, which will be described and made available in an upcoming publication.\\

% No other data were produced or analysed during the course of this work.

%%%%%%%%%%%%%%%%%%%% REFERENCES %%%%%%%%%%%%%%%%%%

% The best way to enter references is to use BibTeX: 

\bibliographystyle{mnras}
\bibliography{refs.bib}
%\bibliography{example} % if your bibtex file is called example.bib

%%%%%%%%%%%%%%%%%%%%%%%%%%%%%%%%%%%%%%%%%%%%%%%%%%

%%%%%%%%%%%%%%%%% APPENDICES %%%%%%%%%%%%%%%%%%%%%

\appendix

\section{A justification that interpolation does not affect the results of this paper} \label{sec:app_interp}

To match the high cadence of the HARPS-N solar RVs, we calculate RVs from the SDO/HMI images (see section \ref{sec:sdo/hmi}) every four hours, and interpolate these RVs onto the timestamps of the HARPS-N observations.\\

To assess the impact that interpolation has on the signal, we produce a month-long time series of RVs, generated from every available 720s exposure set of SDO/HMI images from 2017-Jan-01 to 2017-Feb-01. From this high-cadence RV series we produce a sub-sampled RV series with six RV measurements per day, to match the long-baseline series we use in the bulk of this paper. We then interpolate this sub-sampled RV series back onto the timestamps of the original high-cadence time series.\\

We find that the median absolute deviation between the high-cadence and interpolated RV series is 2.1 \si{\centi\metre\per\second}. This is significantly lower than any other level of variability we discus in this paper, justifying that our choice to interpolate the RVs calculated from the SDO/HMI images does not impact the results of this paper.

\section{The effect of sampling noise on the RMS of the inactive-region RVs.} \label{app:sigma_sigma}

In Fig. \ref{fig:ssn_vs_rms}, we see that a variation in the RMS of inactive-region RV sub-series. 
Whilst each RMS is calculated from a relatively large number of observations, we show in Fig. \ref{fig:structure_functions} that the inactive-region RVs become uncorrelated at timescales longer than a few days.
This relatively large timescale, compared to the 50-day baseline of each sub-series, means that the sampling error on the RMS is larger than would be expected by simply considering the number of observations.\\

\begin{figure}
    \centering
    \includegraphics[width=1\linewidth]{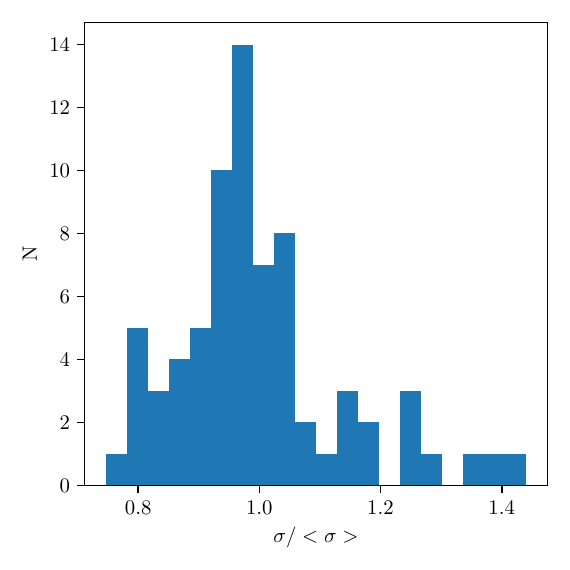}
    \caption{A histogram showing the distribution of RMS values of 50-day-long sub-series of the simulated supergranulation time series of \citet{meunierActivityTimeSeries2019}. Since we are only concerned with the fractional scatter of the RMS measurements, we normalise the x axis by the mean RMS value.}
    \label{fig:sgr_rms_hist}
\end{figure}

To test this, we take the simulated supergranulation time series of \citet{meunierActivityTimeSeries2019} and calculate the RMS for the same 50-day sub-series as in Fig. \ref{fig:ssn_vs_rms}.
This time series is constructed in a stationary manner, so that any variation seen between values of the RMS can with good confidence be attributed to sampling error.
Fig. \ref{fig:sgr_rms_hist} shows the distribution of sub-series RMS values.
As we are only interested in the fractional difference between the individual RMS values, we normalise by the mean RMS.
We see a relative scatter of the RMS values, $\frac{\sigma_\sigma}{<\sigma>}$, of 14 per cent, where 
\begin{equation}
    \frac{\sigma_\sigma}{<\sigma>} = \frac{\sqrt{\frac{1}{n_\sigma}\sum \left( \sigma - \left(\frac{1}{n_\sigma}\sum \sigma\right)\right)^2}}{\frac{1}{n_\sigma}\sum \sigma}.
\end{equation}
This value of $\frac{\sigma_\sigma}{<\sigma>}$ is comparable to the inactive-region $\frac{\sigma_\sigma}{<\sigma>}$ of 17 per cent.
This indicates that the majority of the variation seen in Fig. \ref{fig:ssn_vs_rms} is due to statistical noise.

\section{Structure functions} \label{sec:structure_functions}

% Recently adapted from extra-galactic studies to light curve variability of young stellar objects \citep{sergisonCharacterizingIbandVariability2020, venutiMulticolorVariabilityYoung2021, lakelandUnderstandingYSOVariability2022},
The structure function (SF) is an analysis tool which shows how the variability within a time series changes as a function of timescale. It has had success in analysing photometric time series data \citep[e.g., ][]{hughesUniversityMichiganRadio1992, devriesLongTermVariabilitySloan2003, sergisonCharacterizingIbandVariability2020}. SFs use a pairwise comparison of points in a time series $f(t)$ to calculate the level of variability as a function of timescale. To calculate the variability of points with a given separation $\tau = \left | t_1 - t_2 \right |$, we define the structure function as 

% \begin{equation} \label{eq:structure_function}
% SF(\tau_1, \tau_2) = \frac{1}{2}\frac{1}{N(\tau_1, \tau_2)} \sum \left( f(t_i) - f(t_j) \right) ^2,  
% \end{equation}

\begin{equation} \label{eq:structure_function}
\mathrm{SF}(\tau) =  \langle \left( f(t) - f(t + \tau) \right) ^2\rangle,  
\end{equation}
where the angular brackets indicate an average over all pairs of observations with separation $\tau$. In practice, $\mathrm{SF}$ is calculated for logarithmic bins in $\tau$.
That is to say, in the case of discrete data, the structure function is calculated as
\begin{equation}
    \mathrm{SF}(\tau_1, \tau_2) = \frac{1}{N_\mathrm{p}} \sum_{ij} (f_i - f_j)^2,
\end{equation}
where the summation is taken over all pairs of data points which are separated by $\tau_1 < | t_i - t_j | < \tau_2$  and $N_\mathrm{p}$ is the number of such pairs.\\

\begin{figure}
    \centering
    \includegraphics[width=\linewidth]{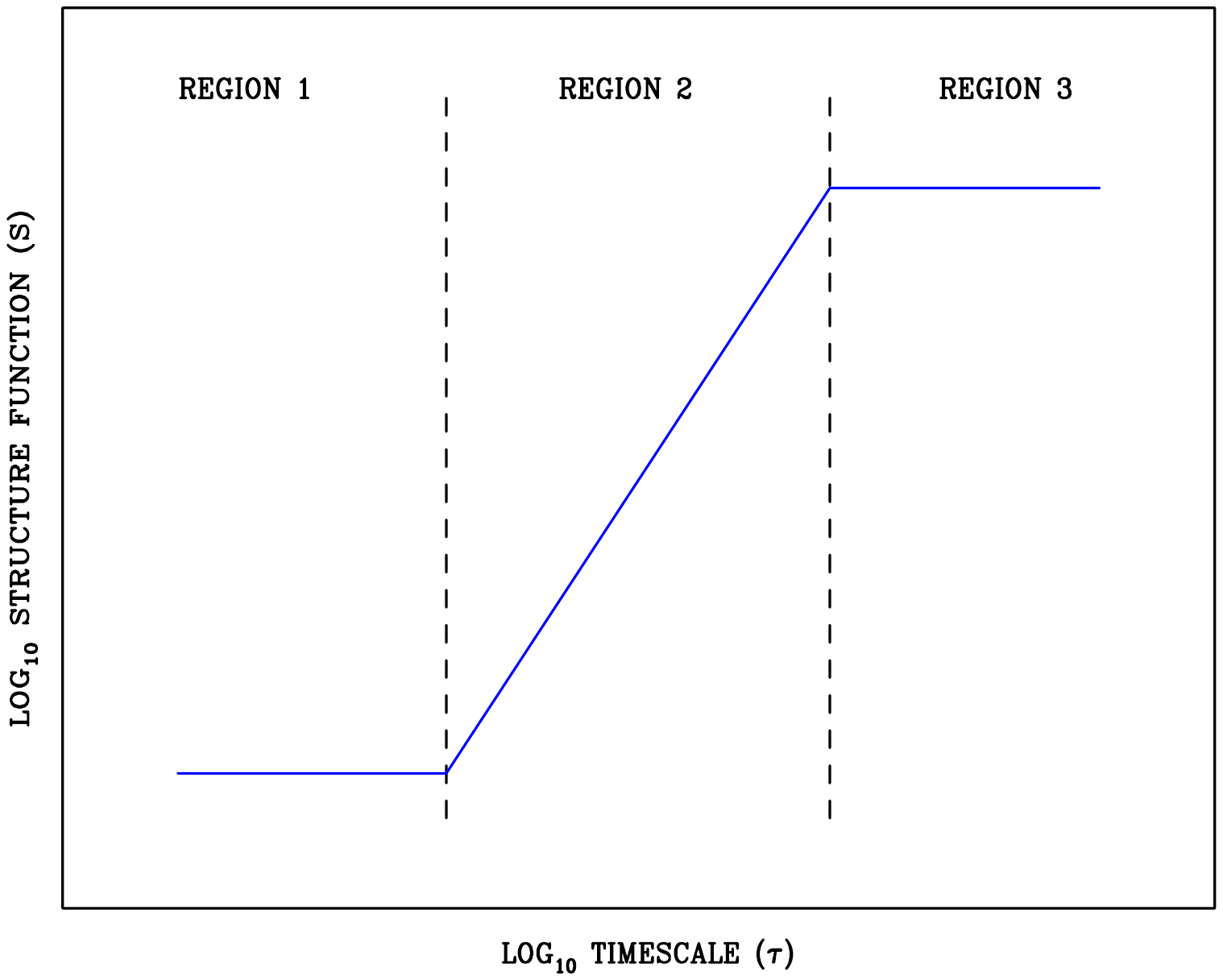}
    \caption{An illustration of the typical properties of structure functions, reproduced from \citet{sergisonCharacterizingIbandVariability2020}. Region 1 shows white noise present at short timescales, often interpreted as instrumental noise. Region 2 shows a typical power-law relationship between timescale and level of variability. Region 3 shows timescales at which there is no further increase in variability. The transition between Regions 2 and 3 shows the characteristic timescale of the variability.}
    \label{fig:sergison_plot}
\end{figure}

Fig. 7 of \citet{sergisonCharacterizingIbandVariability2020}, reproduced here as Fig. \ref{fig:sergison_plot}, demonstrates the typical form of a SF. The short timescale plateau shown in Region 1 indicates any intrinsic variability in the signal is below the noise floor, so the variability is dominated by uncorrelated instrumental noise. In Region 2, there is a power-law increase in the structure function, the gradient of which is determined by the spectrum of variability being investigated. Region 3 shows timescales longer than the dominant variability timescale in the data. Therefore, after this timescale has been reached, the value of the structure function stops increasing, again becoming uncorrelated.
% In Region 3, we are probing timescales longer than the intrinsic timescale of the variability process in the signal, therefore increasing the timescale yields no increase in variability. 
The transition between Regions 2 and 3, referred to as the \lq knee\rq, tells us the characteristic variability timescale.
% Of particular interest are regions 2 and 3 which, respectively, show timescales up to and above the intrinsic timescale of variability. By consequence therefore, the transition between the characteristic power law of region 2 and the flat-spectrum region 3, referred to as the \lq knee\rq , tells us the characteristic timescale of a time series. 
Fig. 5 of \citet{lakelandUnderstandingYSOVariability2022} demonstrates that, for a sinusoidal signal, the knee is at roughly a quarter of the period. This highlights that the timescale identified by a structure function, whilst related to any period present in a signal, is not the same as a period. Another feature of Fig. \ref{fig:sergison_plot} is that the structure function becomes larger as we move towards longer timescales. Whilst not strictly a cumulative measure, this behaviour is typical for most types of variability.\footnote{A notable exception is for periodic or quasi-periodic signals. For such signals, pairs of data points separated by multiples of the period will show little if any variation, causing characteristic dips in the structure function.} This property highlights how short-timescale variations still impact the overall level of variability at longer timescales.\\

An advantage of the structure function over other analysis tools is the direct relationship between the structure function and the level of variability at a given timescale. We show in Appendix \ref{sec:app_sf_rms} that, for an uncorrelated signal, the structure function will tend to the value $2\sigma^2$, where $\sigma$ is the standard deviation of the data. This clear relationship between structure function and standard deviation allows a much more direct interpretation of the level of variability within a signal for a given timescale than, say, a Fourier power spectrum.\\

% Note that the definition of the structure function in Eq. \ref{eq:structure_function} differs from previous definitions by a factor of $\frac{1}{2} $\citep[e.g., ][]{lakelandUnderstandingYSOVariability2022}, in order to draw better comparison with the standard deviation.\\
% We use a slightly different form of the structure function to previous analyses in order to better draw comparison to the standard deviation.\\

For a more detailed description of some properties of structure functions, see section 5 of \citet{lakelandUnderstandingYSOVariability2022} and references therein. To avoid spurious results arising from bins with too few points, we require at least 100 pairs of observations in each $\tau$ bin throughout this paper.\\

\subsection{The relationship between the structure function and standard deviation} \label{sec:app_sf_rms}

Consider an infinite, uncorrelated time series $f(t)$. This time series will have mean 
\begin{equation}
    \mu = \langle f(t) \rangle, 
\end{equation}
and standard deviation,
\begin{equation}
    \sigma = \langle \left ( f(t) - \mu \right )^2 \rangle. 
\end{equation}
Since $f(t)$ is uncorrelated, we can also say that 
\begin{equation}\label{eq:f_autocorr}
    \langle f (t) \times f ( t + x) \rangle = \mu^2 + \sigma^2\delta(x),
\end{equation}
where $\delta{(x)}$ is the $\delta$-function.
% For a particular timescale $\tau$, the structure function $\mathrm{SF}(\tau)$ is given by
% \begin{equation}
%     \mathrm{SF}(\tau) = \frac{1}{2}\langle \left ( f(t) - f(t+\tau) \right)\rangle,
% \end{equation}
% i.e., the infinite limit of Equation \ref{eq:structure_function}. 
From Eq. \ref{eq:structure_function}, it follows that
\begin{equation}
    \mathrm{SF}(\tau) = \langle \left( \left(f (t)  - \mu\right) - \left ( f(t+\tau) - \mu \right)\right) ^2 \rangle 
\end{equation}
\begin{equation}
\begin{split}
    =  &\left[ \langle \left( f(t) - \mu \right) ^ 2\rangle +  \langle \left( f(t + \tau) - \mu \right) ^ 2\rangle \right ] \\
    & \qquad \qquad -  2\langle \left(f(t) - \mu \right ) \times \left( f(t + \tau) - \mu \right) \rangle
\end{split}
\end{equation}
\begin{equation}
    = 2\sigma ^2 - 2\left[ \langle f (t) \times f ( t + x) \rangle -  \mu \langle f(t) \rangle-  \mu \langle f(t + \tau) \rangle   + \mu^2\right].
\end{equation}
From Equation \ref{eq:f_autocorr}, therefore, we see that, for uncorrelated signals,
\begin{equation}
    \mathrm{SF}(\tau) =  2\sigma^2.
\end{equation}
This is equivalent to eq. A4 of \citet{1985ApJ...296...46S}.

\subsection{Estimating the effect of sampling noise} \label{sec:app_sf_err}

As discussed in appendix C of \citet{sergisonCharacterizingIbandVariability2020}, it is non-trivial to estimate the effect of sampling noise on structure function measurements (as we can think of $f(t)$ being drawn at random from an underlying distribution).
The difficulty arises in calculating the number of independent pairs of data points contributing to each measurement.\\

To estimate the level of sampling noise, we divide the simulated supergranulation RV time series of \citet{meunierActivityTimeSeries2019} (see Section \ref{sec:sgr_as_rv_cause}) into year-long sub-series, and calculate the structure function of each.
This is in analogy to fig. 10 of \citet{sergisonCharacterizingIbandVariability2020}.
We opt to use the supergranulation time series as it will allow us to estimate the sampling noise at timescales below and above the characteristic timescale of the variability \citep[as opposed to generating red-noise series as done by][]{sergisonCharacterizingIbandVariability2020} and the time series is constructed to be stationary so we can be confident that any variation between the structure functions is due to the sampling noise.\\

\begin{figure}
    \centering
    \includegraphics[width=1\linewidth]{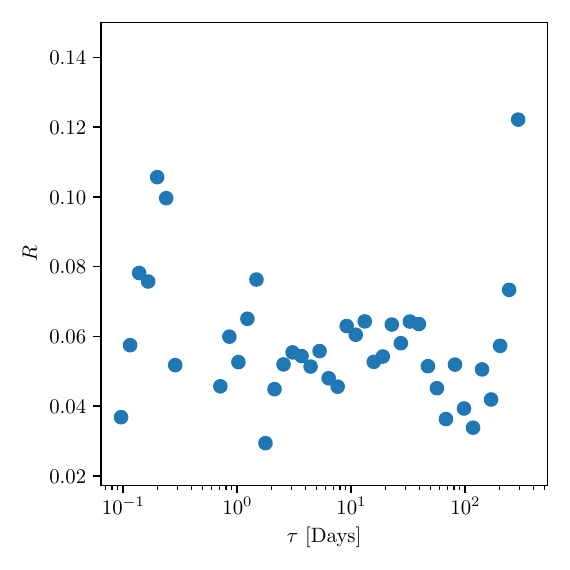}
    \caption{Ratio $R$ of $\sigma_\mathrm{SF}$ to the median SF for 10 independent sub-series of the supergranulation RV time series discussed in Section \ref{sec:sgr_as_rv_cause}.}
    \label{fig:sgr_noise}
\end{figure}

Fig. \ref{fig:sgr_noise} shows the ratio $R$ of standard deviation of $\sqrt{\mathrm{SF}}$ to the median value of $\sqrt{\mathrm{SF}}$  at each timescale.
We show that the typical sampling error is on order of 5 per cent of the median value of $\sqrt{\mathrm{SF}}$ at a given timescale.
Given that these sub-series are drawn from the same distribution, and that they are well-sampled, we expect this to be a lower bound on the uncertainty associated with a structure function measurement.
\\

%%%%%%%%%%%%%%%%%%%%%%%%%%%%%%%%%%%%%%%%%%%%%%%%%%

% Don't change these lines
\bsp	% typesetting comment
\label{lastpage}
\end{document}